\documentclass[modern]{rnaastex}

\newcommand{\msun}{M$_{\sun}$}

\newcommand{\namesh}{2MASS~J1450$-$7841}

\begin{document}

\title{A Candidate Wide Brown Dwarf Binary in the Argus Association: 2MASS~J14504216$-$7841413 and 2MASS~J14504113$-$7841383}

\correspondingauthor{Adam J.\ Burgasser}
\email{aburgasser@ucsd.edu}

\author[0000-0002-6523-9536]{Adam J.\ Burgasser}
\affiliation{Center for Astrophysics and Space Science, University of California San Diego, La Jolla, CA 92093, USA }

\author{Dagny L.\ Looper}
\affiliation{New York University Tisch School of the Arts, 721 Broadway, 10th Floor, New York, NY 10003, USA }

\author{J.\ Davy Kirkpatrick}
\affiliation{IPAC, Mail Code 100-22, Caltech, 1200 E. California Blvd. Pasadena, CA 91125.}


\section{} 

Widely-separated ($\gtrsim$100~AU) multiples are rare
among the lowest mass stars and brown dwarfs \citep{2007ApJ...667..520C,2009ApJ...703.1511K}, 
and often (but not exclusively) associated with 
young ($\lesssim$100~Myr), nearby stellar associations (e.g., \citealt{2007ApJ...660.1492C}).
We report the discovery of a
wide, very low mass, and potentially young binary,
2MASS~J14504216$-$7841413 and 2MASS~J14504113$-$7841383 (hereafter {\namesh}AB).
The primary was initially identified in the DENIS \citep{1997Msngr..87...27E} 
and 2MASS \citep{2006AJ....131.1163S} surveys
as a candidate late-M dwarf based on its red optical/infrared colors
(DENIS $I$ = 18.03$\pm$0.16, $I-J$ = 3.32$\pm$0.19,
2MASS $J$ = 14.71$\pm$0.06, $J-K_s$  = 1.38$\pm$0.08).
A faint red companion, 2MASS~J14504113$-$7841383, is present in the 2MASS Point Source Catalog 
($J$ = 15.64$\pm$0.09, $J-K_s$  = 1.28$\pm$0.14)
separated by 4$\farcs$23$\pm$0$\farcs$11 at position angle 314$\degr$.

\begin{figure}[h!]
\begin{center}
\includegraphics[trim={1cm 3cm 1cm 4.5cm},clip,scale=0.7,angle=0]{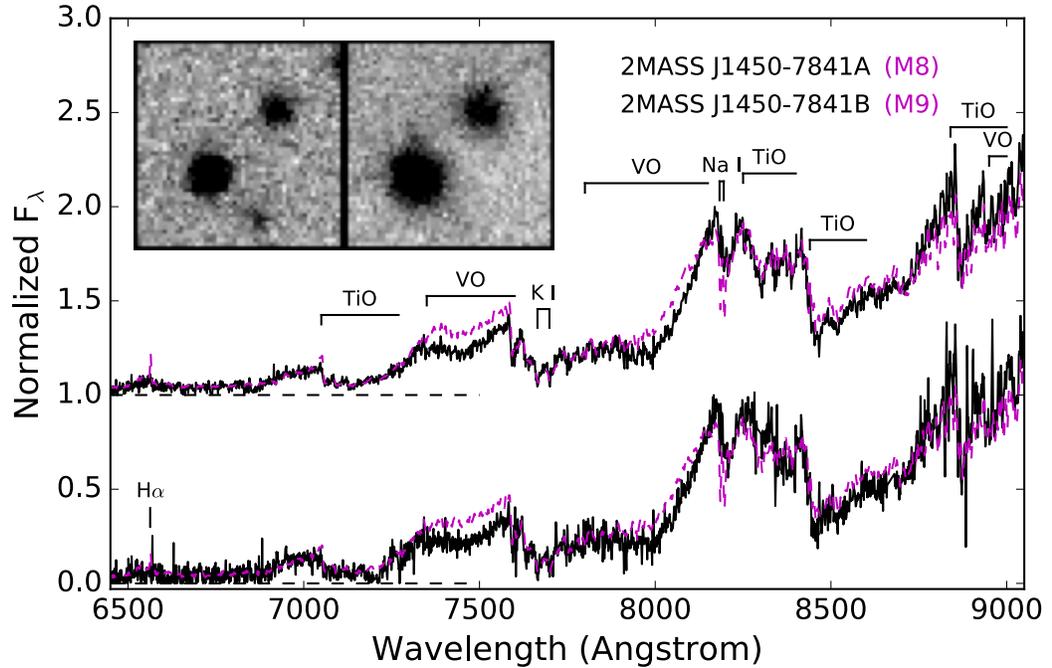}
\caption{LDSS-3 spectra of {\namesh}A (top) and B (bottom), corrected for telluric absorption and normalized at 8170~{\AA}; compared to M8 and M9 templates (dashed magenta lines). Key absorption features are labeled. Inset shows LDSS-3 $i$-band (left) and $z$-band (right)
images of {\namesh}AB, 9$\farcs$5 on a side and oriented with North up and East to the left. {\namesh}A is the southeastern component. \label{fig:1}}
\end{center}
\end{figure}

We observed {\namesh}AB with the Low Dispersion Survey Spectrograph 3 (LDSS-3; \citealt{1994PASP..106..983A})
on the Magellan 6.5m Clay Telescope on 2007 May 9 (UT). The system was resolved in $i$- and $z$-band imaging data at an identical separation (4$\farcs$28$\pm$0$\farcs$14 at 314$\degr$; Figure~\ref{fig:1}) as 2MASS, with $\Delta{i}$ = 1.03$\pm$0.05 and $\Delta{z}$ = 1.01$\pm$0.05.  
Combining astrometry from 2MASS and LDSS-3, we measure identical proper motions of ($\mu_{\alpha}\cos{\delta}$,$\mu_{\delta}$) = (22$\pm$32,$-$59$\pm$34)~mas/yr for {\namesh}A and 
 ($-$12$\pm$23,$-$4$\pm$23)~mas/yr for {\namesh}B.
 
LDSS-3 red optical spectroscopy (Figure~\ref{fig:1}; cf.~\citealt{2006ApJ...645.1485B})
reveals strong TiO and VO absorption bands in both sources, typical of late-M dwarfs. Comparing to spectral templates
\citep{2007AJ....133..531B}, we estimate types of M8p and M9p for {\namesh}A and B, respectively. The ``peculiar'' designations reflect the strong VO bands and weak Na~I lines, consistent with low surface gravities \citep{2008ApJ...689.1295K,2009AJ....137.3345C}. We detect weak H$\alpha$ emission from {\namesh}B, but the data are inadequate to infer the presence of Li~I in either spectrum.

Using the young and field $M_J$/spectral type relations of \citet{2016ApJS..225...10F}, we estimate equivalent distances of 75$\pm$25~pc for {\namesh}A and B, implying a projected separation of $\sim$300~AU. 
Applying this distance estimate and the proper motions to the BANYAN II tool \citep{2014ApJ...783..121G,2013ApJ...762...88M}, we find a 93\% probability of membership in the $\sim$40~Myr Argus
Association \citep{2011ApJ...732...61Z}. At this age, {\namesh}A and B have masses of $\sim$0.04~{\msun} \citep{2003A&A...402..701B}, well below the hydrogen fusion limit. 
Confirmation and study of this potentially young and unusually wide substellar binary is warranted.

\acknowledgments

This paper includes data gathered with the 6.5 meter Magellan Telescopes located at Las Campanas Observatory, Chile.

\clearpage

\end{document}